\documentclass[twocolumn]{aastex631}

\newcommand{\hst}{\emph{HST}}
\newcommand{\jwst}{\emph{JWST}}

\usepackage{amsmath}

\shorttitle{Improve photo-$z$ with the Ly$\alpha$ damping wing absorption}
\shortauthors{Asada et al.}
\graphicspath{{./}{figures_v2cats/}}
\begin{document}

\title{Improving photometric redshifts of Epoch of Reionization galaxies:\\ a new empirical transmission curve with neutral hydrogen damping wing Ly$\alpha$ absorption}

\correspondingauthor{Yoshihisa Asada}
\email{asada@kusastro.kyoto-u.ac.jp}

\author[0000-0003-3983-5438]{Yoshihisa Asada}
\affiliation{Department of Astronomy and Physics and Institute for Computational Astrophysics, Saint Mary's University, 923 Robie Street, Halifax, Nova Scotia B3H 3C3, Canada}
\affiliation{Department of Astronomy, Kyoto University,
Sakyo-ku, Kyoto 606-8502, Japan}

%% Direct contributions
\author[0000-0001-8325-1742]{Guillaume Desprez}
\affiliation{Department of Astronomy and Physics and Institute for Computational Astrophysics, Saint Mary's University, 923 Robie Street, Halifax, Nova Scotia B3H 3C3, Canada}
\affiliation{Kapteyn Astronomical Institute, University of Groningen, P.O. Box 800, 9700AV Groningen, The Netherlands}

\author[0000-0002-4201-7367]{Chris J. Willott}
\affiliation{NRC Herzberg, 5071 West Saanich Rd, Victoria, BC V9E 2E7, Canada}

\author[0000-0002-7712-7857]{Marcin Sawicki}
\affiliation{Department of Astronomy and Physics and Institute for Computational Astrophysics, Saint Mary's University, 923 Robie Street, Halifax, Nova Scotia B3H 3C3, Canada}

%% CANUCS founders/builders
%% also includes external collabs
%\author[0000-0002-4542-921X]{Roberto Abraham}
%\affiliation{David A. Dunlap Department of Astronomy and Astrophysics, University of Toronto, 50 St. George Street, Toronto, Ontario, M5S 3H4, Canada}

\author[0000-0001-5984-0395]{Maru{\v s}a Brada{\v c}}
\affiliation{Faculty of Mathematics and
Physics, University of Ljubljana, Jadranska ulica 19, SI-1000 Ljubljana, Slovenia}
\affiliation{Department of Physics and Astronomy, University of California Davis, 1 Shields Avenue, Davis, CA 95616, USA}

\author[0000-0003-2680-005X]{Gabriel Brammer}
\affiliation{Cosmic Dawn Center (DAWN), Denmark}
\affiliation{Niels Bohr Institute, University of Copenhagen, Jagtvej 128, DK-2200 Copenhagen N, Denmark}

%\author[0000-0001-8489-2349]{Vince Estrada-Carpenter}
%\affiliation{Department of Astronomy and Physics and Institute for Computational Astrophysics, Saint Mary's University, 923 Robie Street, Halifax, Nova Scotia B3H 3C3, Canada}

\author[0000-0002-6533-2810]{Florian Dubath}
\affiliation{Department of Astronomy, University of Geneva, ch. d'Ecogia 16, 1290 Versoix, Switzerland}

\author[0000-0001-9298-3523]{Kartheik G. Iyer}
\affiliation{Columbia Astrophysics Laboratory, Columbia University, 550 West 120th Street, New York, NY 10027, USA}

\author[0000-0003-3243-9969]{Nicholas S. Martis}
%\affiliation{NRC Herzberg, 5071 West Saanich Rd, Victoria, BC V9E 2E7, Canada}
%\affiliation{Department of Astronomy and Physics and Institute for Computational Astrophysics, Saint Mary's University, 923 Robie Street, Halifax, NS B3H 3C3, Canada}
\affiliation{Faculty of Mathematics and
Physics, University of Ljubljana, Jadranska ulica 19, SI-1000 Ljubljana, Slovenia}

%\author[0000-0002-7547-3385]{Jasleen Matharu}
%\affiliation{Cosmic Dawn Center (DAWN), Denmark}
%\affiliation{Niels Bohr Institute, University of Copenhagen, Jagtvej 128, DK-2200 Copenhagen N, Denmark}

%\author[0000-0002-8530-9765]{Lamiya Mowla}
%\affiliation{Whitin Observatory, Department of Physics and Astronomy, Wellesley College, 106 Central Street, Wellesley, MA 02481, USA}

\author[0000-0002-9330-9108]{Adam Muzzin}
\affiliation{Department of Physics and Astronomy, York University, 4700 Keele St. Toronto, Ontario, M3J 1P3, Canada}

\author{Gaël Noirot}
\affiliation{Space Telescope Science Institute, 3700 San Martin Drive, Baltimore, Maryland 21218, USA}
\affiliation{Department of Astronomy and Physics and Institute for Computational Astrophysics, Saint Mary's University, 923 Robie Street, Halifax, Nova Scotia B3H 3C3, Canada}

\author[0000-0002-8108-9179]{St\'ephane Paltani}
\affiliation{Department of Astronomy, University of Geneva, ch. d'Ecogia 16, 1290 Versoix, Switzerland}

\author[0000-0001-8830-2166]{Ghassan T. E. Sarrouh}
\affiliation{Department of Physics and Astronomy, York University, 4700 Keele St. Toronto, Ontario, M3J 1P3, Canada}

%\author[0000-0002-6338-7295]{Victoria Strait}
%\affiliation{Cosmic Dawn Center (DAWN), Denmark}
%\affiliation{Niels Bohr Institute, University of Copenhagen, Jagtvej 128, DK-2200 Copenhagen N, Denmark}

%\author[0000-0002-9842-6354]{Johannes Zabl}
%\affiliation{Department of Astronomy and Physics and Institute for Computational Astrophysics, Saint Mary's University, 923 Robie Street, Halifax, NS B3H 3C3, Canada}

\author[0000-0001-9414-6382]{Anishya Harshan}
\affiliation{Faculty of Mathematics and
Physics, University of Ljubljana, Jadranska ulica 19, SI-1000 Ljubljana, Slovenia}

\author[0000-0002-5694-6124]{Vladan Markov}
\affiliation{Faculty of Mathematics and
Physics, University of Ljubljana, Jadranska ulica 19, SI-1000 Ljubljana, Slovenia}

%\collaboration{20}{CANUCS}



\begin{abstract}
We present a new analytical model for the attenuation to Epoch of Reionization (EoR) galaxies by proximate neutral hydrogen gas.
Many galaxy spectra in the EoR taken by \jwst\ have shown a flux deficit at wavelengths just redward of the Lyman break, and this has been regarded as resulting from Ly$\alpha$ damping wing absorption by the increasing amount of neutral hydrogen in the line-of-sight. 
However, previous attenuation models for the intergalactic medium (IGM) commonly used in photometric redshift template-fitting codes assume that the Lyman break is rather sharp, which leads to systematic overestimation of photometric redshifts at $z>7$.
In this letter, we build and empirically calibrate a new attenuation model that takes the increased Ly$\alpha$ damping wing absorption into account.
%, and empirically calibrate the model in the EoR with \jwst\ observations from the CANUCS program.
Our model consists of the canonical IGM absorption and an additional absorption component due to dense neutral hydrogen gas clouds proximate to the galaxy, and we derive the redshift evolution of 
% the CGM 
H{\sc i} column density of the proximate clouds by calibrating the model using CANUCS \jwst\ observations.
The resulting total transmission curve resolves the photometric redshift bias at $z>7$, an improvement that is robust to choice of template-fitting code, template set, and photometric catalog used. The new attenuation model can be easily implemented in existing template-fitting codes, and significantly improves the photometric redshift performance in the EoR.

\end{abstract}



\section{Introduction} \label{sec:intro}
Redshift estimations of galaxies in the distant universe based on their photometry, namely photometric redshifts (photo-$z$), have been an essential tool in extragalactic studies \citep[e.g.,][]{Sawicki1997, FernandezSoto1999, Sawicki2002, Ilbert2009, Muzzin2013, Tanaka2017, Moutard2020, Weaver2022}.
One of the key photometric signatures used for estimating the redshifts of high-$z$ galaxies is the break imprinted in the spectral energy distributions (SEDs) of distant galaxies at $\lambda_{\rm rest}=1216$ \AA\ and commonly termed ``Lyman break" in recent high-$z$ literature. Its observational effect is to produce a strong photometric `drop-out' between filters that bracket rest-frame 1216\AA, similar in its effect to the 912\AA\ break that produces $u$-band and $g$-band dropouts at lower redshifts \citep[e.g.,][]{Guhathakurta1990, Steidel1996ApJ, Steidel1999}. Both the 912\AA\ and 1216\AA\ breaks are caused by neutral hydrogen foreground to the source, although the exact mechanisms are slightly different: the rest-frame 912\AA\ Lyman continuum break is caused by the $n=1 \rightarrow \infty$ photoionization and is the dominant effect for lower redshift objects ($z\sim3-4$), while the 1216\AA\ Lyman-$\alpha$ break is due to resonant scattering and absorption by the $n=1 \rightarrow 2$ bound-bound transition and dominates at $z\gtrsim5$. Both breaks are commonly referred to as ``Lyman break" in the literature and have been used to efficiently select photometric `drop-out' galaxies and/or estimate their redshift solely from photometry, including identifying galaxy candidates up to $z>14$ \citep{Hainline2024ApJ,Robertson2024}. In this Letter we primarily focus on the imprint of the rest-frame 1216~\AA\ break on $z>6$ galaxy photometry.

%Lyman-$\alpha$ break at $\lambda_{\rm rest}=1216$ \AA, a.k.a. the Lyman break, that is due to the  absorption of high-$z$ galaxy photons by neutral hydrogen along the line of sight \citep[Lyman break selection; e.g.,][]{Steidel1996ApJ}\footnote{Originally in \citet{Steidel1996ApJ}, the Lyman break is defined as the break at $\lambda_{\rm rest}=912$ \AA\ that is relevant at $z\sim3$. However, in this paper, we define the "Lyman break" as the break at $\lambda_{\rm rest}=1216$ \AA\ that is more relevant at higher redshifts ($z\gtrsim5$).}.

%The Lyman break method 
%has long been used to efficiently select high-$z$ galaxies and/or estimate their redshifts solely from photometry, and the method has been shown successful a

Since the launch of \jwst, it has been relatively easy to spectroscopically confirm the redshifts of the high-$z$ candidates selected by their (rest-frame 1216\AA) Lyman break signature. %, although accurate photometric redshifts ramain vital in this field of research to both select subsamples for spectroscopy and to provide large samples for statistica studies. .
Recent work with \jwst\ has shown high success rates of identifying high-$z$ galaxies based on photo-$z$, however, many authors have consistently reported that photo-$z$ at $z>7$ are systematically overestimated by $\delta z (= z_{\rm phot} - z_{\rm spec})\sim$0.2 compared to spectroscopy  \citep[e.g.,][]{Arrabal_Haro2023ApJ,Fujimoto2023ApJ,Helton2023arXiv,Finkelstein2024ApJ,Hainline2024ApJ,Willott2024ApJ}.

%This overestimation is likely due to the growing Ly$\alpha$ damping wing absorption at $z>7$; the Lyman break at $\lambda_{\rm rest}=1216$ \AA\ has been assumed to be a sharp break in the canonical IGM transmission models \citep[e.g.,][]{Madau1995ApJ,Inoue2014mn}, while a large number of \jwst/NIRSpec observations have found that the spectral break is rather smooth and spectra typically show a flux deficit in the wavelength range just redward of the Ly$\alpha$ line \citep[e.g.,][]{Umeda2023arXiv,Keating2024MNRAS}.
%The damped absorption has been interpreted as the result of the increasing amount of neutral hydrogen gas in the $z>7$ galaxy line-of-sight, related to cosmic reionization at this epoch \citep[e.g.,][]{Curtis-Lake2023NatAs}.
%Therefore, properly characterizing the line-of-sight neutral hydrogen absorption around the Lyman break of typical galaxies in the Epoch of Reionization (EoR) is essential to improve their photo-$z$ estimations, which leads to reducing the systematic uncertainties of all basic physical properties of EoR galaxies such as luminosity, stellar masses, or star formation rates.

This bias in photometric redshift measurement is likely due to increased Ly$\alpha$ damping wing absorption in the EoR. Canonical intergalactic medium (IGM) transmission models \citep[e.g.,][]{Madau1995ApJ,Inoue2014mn} commonly used by photo-$z$ codes have a sharp break at the rest-frame Ly$\alpha$ wavelength. However,  \jwst/NIRSpec observations of $z>7$ galaxies have typically found that the spectral break is smoother with a flux deficit in the wavelength range just redward of the Ly$\alpha$ line \citep[e.g.,][]{Curtis-Lake2023NatAs,Heintz2024Sci,Umeda2023arXiv}. This smoother spectral shape is partly due to residual neutral hydrogen within ionized bubbles close to galaxies \citep{Keating2024MNRAS}. However, in many cases, additional absorption from proximate damped Ly$\alpha$ systems is also required to fit the spectra \citep{Carniani2024,D'Eugenio2024A&A,Hainline2024ApJ_b,Heintz2024Sci,Hsiao2024ApJ,Umeda2023arXiv,Witstok2024arXiv}. 
To improve photo-$z$ estimations for galaxies in the EoR, it is crucial to characterize the line-of-sight neutral hydrogen absorption around the Lyman break of typical high-$z$ galaxies.
Such improvements in photo-$z$s will then naturally reduce the systematic biases of all basic physical properties of EoR galaxies such as luminosity, stellar masses, or star formation rates. In certain redshift ranges, such improvements can be especially critical as even small overestimate in redshifts will result in emission line contributions to photometry being improperly accounted for. 

In this Letter, we present a new analytical model for the attenuation by line-of-sight neutral hydrogen, including the Ly$\alpha$ damping wing absorption in the EoR, that leads to significantly reduced biases in the photometric redshifts of $z>7$ galaxies. Leveraging the large \jwst/NIRCam and NIRSpec spectroscopic sample of high-$z$ galaxies from the CAnadian NIRISS Unbiased Cluster Survey \citep[CANUCS;][]{Willott2022PASP}, we derive the redshift evolution of the Ly$\alpha$ damping wing absorption imprinted in typical galaxy SEDs at $z>6$. Using our new model of the evolving line-of-sight attenuation leads to improvements in the photo-$z$ estimation in the EoR that essentially eliminates the systematic biases present in previous studies. 

%We utilize \jwst/NIRCam and NIRSpec observations from the CAnadian NIRISS Unbiased Cluster Survey \citep[CANUCS;][]{Willott2022PASP} of a large sample of spectroscopically confirmed $z>6$ galaxies distributed across five different sight lines of the sky. Leveraging the unique sample of high-$z$ galaxies,  we derive the redshift evolution of the Ly$\alpha$ damping wing absorption imprinted in typical galaxy SEDs at $z>6$, to properly model the line-of-sight neutral hydrogen attenuation and to improve the photo-$z$ estimations in the EoR.

Although most of the work in this paper is cosmology-independent, we assume a flat $\Lambda$CDM cosmology with $\Omega_\Lambda=0.7,\ \Omega_m=0.3$, and $H_0=70\ {\rm km\ s^{-1}\ Mpc^{-1}}$ when setting priors for photo-$z$ fitting.

\section{Data}\label{sec:data}
%We use the CANUCS observations to calibrate our model for the neutral hydrogen absorption imprinted in high-$z$ galaxies' SEDs (see Section \ref{subsec:model_DLA} for the details of the model).

This work uses JWST photometry and spectra taken by the CAnadian NIRISS Unbiased Cluster Survey (CANUCS). Full details of the survey, including imaging reductions and photometry will be presented in a future paper (Sarrouh et al. in prep.). Here we briefly present the CANUCS dataset used in this work.
Note that a brief overview of the CANUCS dataset can be found in \citet{Willott2024ApJ}.

The CANUCS program targets five strong lensing cluster fields (A370, MACS0416, MACS0417, MACS1149, and MACS1423).
Each cluster has several pointings, but we use two of them: the central cluster fields (CLU), and the  NIRCam flanking field (NCF).
The CLU fields were observed with both NIRCam and NIRISS, while the NCF fields were observed with only NIRCam.
The CANUCS program also carried out NIRSpec {\tt Prism} follow-up observations with the Micro-Shutter Assembly with a typical exposure time of $\sim3$ ks.
The follow-up targets were selected to cover various science cases, based on the photometric catalogs from Cycle 1 NIRCam and NIRISS observations supplemented by archival \hst\ optical observations.
The NIRSpec follow-up targets were selected from the five CLU fields and one NCF field (MACS0417).
In this work, we use a sample of spectroscopically confirmed galaxies in the EoR (see Section \ref{subsec:meas_DLA} for the sample selection), thus we use NIRCam and NIRISS imaging observations in the five CLU fields and NIRCam observations in the MACS0417 NCF field for photometry.
We also leverage additional supporting imaging data when available: \hst\ optical imaging observations in all six fields (mostly from HST-GO-13504 PI Lotz and HST-GO-16667 PI Brada\v{c}), and NIRISS F090W imaging observations in three CLU fields (A370, MACS0416, and MACS1149) from a \jwst\ Cycle 2 GO program (GO-3362 PI Muzzin).
All available filters and typical depths in each field can be found in \citet{Willott2024ApJ,Desprez2024MNRAS}.
%Consequently, CLU fields have 8 NIRCam filters and 3 NIRISS filters (+1 only in A370 and MACS0416): NIRCam/F090W, F115W, F150W, F200W, F356W, F410M, and F444W, and NIRISS/F115W, F150W, and F200W (+F090W), with supplemented by \hst/ACS F435W, F606W, and F814W.
%The MACS0417 NCF field has 14 NIRCam filters: F090W, F115W, F140M, F150W, F162M, F182M, F210M, F250M, F277W, F300M, F335M, F360M, F410M, and F444W, with supplemented by \hst/WFC3 UVIS F438W and F606W.

The imaging data processing is similar to \citet{Noirot2023MNRAS}, and photometry is as described by \citet{Asada2024MNRAS}.
Briefly, we create a $\chi_{\rm mean}$-detection image \citep[see, e.g.,][]{Drlica_Wagner2018ApJS} from all available \hst\ optical, \jwst/NIRCam, and NIRISS images in each field after removing the bright cluster galaxies \citep[see][for the bCG removal process]{Martis2024arXiv}.
Source detection and deblending are done on the $\chi_{\rm mean}$-detection images with the {\tt photutils} package.
In this work, we use 0.3$^{\prime\prime}$-diameter aperture photometry performed on PSF-homogenized images (matched to F444W PSF) of all filters \citep[see][for the PSF construction and homogenization]{Sarrouh2024}.

The NIRSpec data processing is done in the same way as in \citet{Desprez2024MNRAS}. The NIRSpec data are processed with a combination of the STScI \jwst\ pipeline and the {\tt msaexp} package \citep{Brammer2022zndo}, to obtain the background-subtracted 2D spectrum for each source.
We then extract the 1D spectrum following the "optimal" extraction described in \citet{Horne1986PASP}.
For each 1D spectrum, we run spline fitting with the {\tt msaexp} package as a first pass on the redshift measurements, and then perform visual inspection of all spectra and fits to assign a spectroscopic redshift (spec-$z$) quality grade.

\section{Method}\label{sec:method}
\subsection{Modeling the Ly$\alpha$ damping wing absorption}\label{subsec:model_DLA}

A downturn of the continuum in the spectral region just redward of the Ly$\alpha$ line can be due to the highly neutral IGM \citep[e.g.,][]{Miralda1998ApJ,Curtis-Lake2023NatAs}, dense neutral hydrogen gas near the galaxy \citep[e.g.,][]{Heintz2024Sci}, and/or two-photon process of the nebular continuum emission \citep[e.g.,][]{Cameron2024mn,Mowla2024arXiv}, and these factors are largely degenerate in photometric data. However, rather than trying to disentangle the contributions of these possible effects to smoothed shape of the break, the goal of this paper is to produce a simple, empirically-calibrated model that effectively captures the net effect of the Ly$\alpha$ damping wing absorption on typical SEDs of high-$z$ galaxies, so that public template-fitting codes can use it across all redshifts and improve photo-$z$ estimations at high-$z$.

%We thus assume a canonical IGM transmission curve as a function of redshift \citep[e.g.,][]{Madau1995ApJ,Inoue2014mn} that is commonly used in many template-fitting codes ({\bf REF}), and model the damped Ly$\alpha$ absorption feature on top of this IGM transmission by considering the presence of dense H{\sc i} gas clouds locally associated with the host galaxy (CGM H{\sc i}). We here highlight that modeling the damping wing absorption component independently of the IGM transmission makes it easier and computationally faster to implement the additional damping wing absorption in commonly used template-fitting codes, and also minimizes the effect on the photo-$z$ estimations at low-$z$.

Our empirical model uses a canonical, redshift-dependent IGM transmission curve
\citep[e.g.,][]{Madau1995ApJ,Inoue2014mn} as commonly used in many template-fitting codes on top of which we add an empirically-calibrated Ly$\alpha$ damping wing absorption.  We model the latter by assuming that it arises from the presence of dense H{\sc i} gas clouds locally associated with the host galaxy and we call it Circumgalactic Medium (CGM) H{\sc i} in this paper for convenience, but we want to emphasize that the actual physical origin of this absorption has no bearing on the photo-$z$ estimation. As an additional benefit, modeling the damping wing absorption component independently of the IGM transmission makes it easier to implement and computationally faster in commonly used template-fitting codes, while also minimizing the effect on photo-$z$ estimations at lower redshifts.

Combining the intervening H{\sc i} gas absorption from the line-of-sight IGM and local CGM H{\sc i}, the net optical depth of the IGM+CGM H{\sc i} gas can be written as $\tau_{\rm tot}(\lambda_{\rm obs}, z_s) = \tau_{\rm IGM}(\lambda_{\rm obs}, z_s) + \tau_{\rm CGM}(\lambda_{\rm obs}, z_s)$, where $z_s$ is the the source redshift and $\lambda_{\rm obs}$ is the observed-frame wavelength.
We use the canonical IGM transmission curve of \citet{Inoue2014mn} as $\tau_{\rm IGM}(\lambda_{\rm obs}, z_s)$.
Following the formulation in \citet{Totani2006pasj}, the Ly$\alpha$ damping absorption feature due to the CGM is modeled as $\tau_{\rm CGM}(\lambda_{\rm obs}, z_s) = N_{\rm H\textsc{i}, CGM} \sigma_\alpha[\nu_{\rm obs}(1+z_s)]$, where $\nu_{\rm obs}=c/\lambda_{\rm obs}$ is the observed frequency, $N_{\rm H\textsc{i}, CGM}$ is the H{\sc i} gas column density of the CGM, and $\sigma_\alpha(\nu)$ is the Ly$\alpha$ cross section at the rest-frame frequency $\nu$.
The Ly$\alpha$ cross section is defined as
\begin{equation}
    \sigma_{\alpha}(\nu) = \frac{3\lambda_\alpha^2f_\alpha\Lambda_{\rm cl,\alpha}}{8\pi}\frac{\Lambda_\alpha(\nu/\nu_\alpha)^4}{4\pi^2(\nu - \nu_\alpha)^2 + \Lambda_\alpha^2(\nu/\nu_\alpha)^6/4},
\end{equation}
where $f_\alpha=0.4162$ is the absorption oscillator strength, and $\Lambda_\alpha=3(g_u/g_l)^{-1}f_\alpha\Lambda_{\rm cl, \alpha}$ is the damping constant of the Ly$\alpha$ resonance.
The $\nu_\alpha$ and $\lambda_\alpha$ are the rest-frame Ly$\alpha$ transition frequency and wavelength, respectively.
The $g_u$ and $g_l$ factors are the statistical weights of the upper and lower levels, and $g_u/g_l=3$ for the Ly$\alpha$ transition.
The classical damping constant is $\Lambda_{\rm cl,\alpha} = (8\pi^2e^2)/(3m_ec\lambda_\alpha^2)=1.503\times10^9\ {\rm s^{-1}}$ in the CGS unit system, where $e$ and $m_e$ are the electron charge and mass, respectively.

Based on the formulation above, the CGM H{\sc i} gas optical depth $\tau_{\rm CGM}(\lambda_{\rm obs}, z_s)$ can be written as a function of the CGM {\sc Hi} column density $N_{\rm H\textsc{i}, CGM}$, the source redshift $z_s$, and the observed wavelength $\lambda_{\rm obs}$.
In the following, we explore the typical CGM {\sc Hi} column density $N_{\rm H\textsc{i}, CGM}$ at $z>6$ as a function of the source redshift $z_s$ so that the optical depth $\tau_{\rm CGM}(\lambda_{\rm obs}, z_s)$ at a given redshift can be determined as a function of the observed wavelength $\lambda_{\rm obs}$.

\subsection{Measuring $N_{\rm H\textsc{i}, CGM}$ at $z>6$}\label{subsec:meas_DLA}
Our goal is to provide a simple empirically-calibrated model that can improve photo-$z$ estimations by modeling the effect of the damping wing absorption on photometric SEDs of high-$z$ galaxies. For this purpose, we use a sample of galaxies selected from the CANUCS observations to meet both of the following criteria:
\begin{enumerate}
    \item Secure spectroscopic redshift measurement at $z_{\rm spec}>6$ with \jwst/NIRSpec observations based on multiple emission lines,
    \item Observed with \jwst/NIRCam and with reliable NIRCam photometry (i.e., not too close to masked regions  or bright neighbors).
\end{enumerate}
We highlight that we do not require the detection of the rest-frame UV continuum and the Lyman break in the NIRSpec spectra, but simply an accurate,  spectroscopic redshift measurement. A total of 102 galaxies are selected, with $\langle z \rangle = 7.12$ and $\langle M_{\rm UV} \rangle = -18.95$ AB mag after the gravitational lensing correction with gravitational lens models by \citet{Gledhill2024ApJ,Rihtarsic2024arXiv}; and Desprez et al. in prep.  
The sample galaxies distribute across the six CANUCS fields, and the median magnification factor is $\langle \mu \rangle = 1.56$.
The sample thus includes a large number of typical sub-$L^\star$ galaxies at this epoch ensuring that the IGM+CGM transmission curve is calibrated for photometric sources typically found with \jwst\ at $z>6$.

We exploit these at $z_{\rm spec}>6$ galaxies to explore the $N_{\rm H\textsc{i}, CGM}$-$z$ relation so that the net optical depth $\tau_{\rm tot}(\lambda_{\rm obs}, z_s) = \tau_{\rm IGM}(\lambda_{\rm obs}, z_s) + \tau_{\rm CGM}(\lambda_{\rm obs}, z_s)$ represents the characteristic absorption feature in galaxy SEDs at each redshift.
For this, we first measure the $N_{\rm H\textsc{i}, CGM}$ value for each sample galaxy that best reproduces the shape of the observed (photometric) SED.
We use the template fitting code {\tt EAzY-py} \citep[][\footnote{\url{https://github.com/gbrammer/eazy-py}}]{Brammer2008ApJ} and derive the best-fitting template spectrum fixing at $z=z_{\rm spec}$.
In this run, we do not consider the presence of CGM H{\sc i} gas absorption and only use classical IGM absorption \citep[][]{Inoue2014mn}.
Using the best-fitting template spectrum at $z=z_{\rm spec}$, we then estimate the CGM H{\sc i} column density that minimizes the $\chi^2$ value by applying additional CGM H{\sc i} absorption on the best-template spectrum.
Here we use only NIRCam or NIRISS photometry, and search for the best-fit $N_{\rm H\textsc{i}, CGM}$ value within the range of $\log_{10}( N_{\rm H\textsc{i}, CGM}/{\rm cm^{-2}})=[18.0, 23.0]$.

In the template spectrum fitting, we use a custom template set developed by the CANUCS program (Sarrouh et al. in prep.). %Further discussion of this custom templates will be found in a separate paper, but we here briefly describe the template set we used.
The template set is composed of the standard templates {\tt tweak\_fsps\_QSF\_12\_v3} from {\tt EAzY} and modified templates based on the {\tt binc100z001age6\_cloudy\_LyaReduced} template from \citet{Larson2023ApJ}.
We generate two templates based on {\tt binc100z001age6\_cloudy\_LyaReduced}, one of which has all emission lines removed, and the other has boosted [O{\sc iii}]4959/5007 lines.
We also increase the number of wavelength grids to assign reasonable emission line widths of ${\rm FWHM}=200\ {\rm km\ s^{-1}}$ for strong rest-optical emission lines (H$\alpha$, H$\beta$, H$\gamma$, H$\delta$, H$\epsilon$, [O{\sc iii}]4959/5007, and [O{\sc ii}]3726/3729).
These modifications improve the photo-$z$ estimations of extreme emission line galaxies that are typically found in the high-$z$ universe \citep[e.g.,][]{Withers2023ApJ,Boyett2024arXiv}, particularly when a number of medium-band filters are available with \jwst/NIRCam. 
We thus use a total of 15 template spectra in the template fitting with {\tt EAzY-py}.

\begin{figure}[tb]
    \centering
\includegraphics[width=0.95\linewidth]{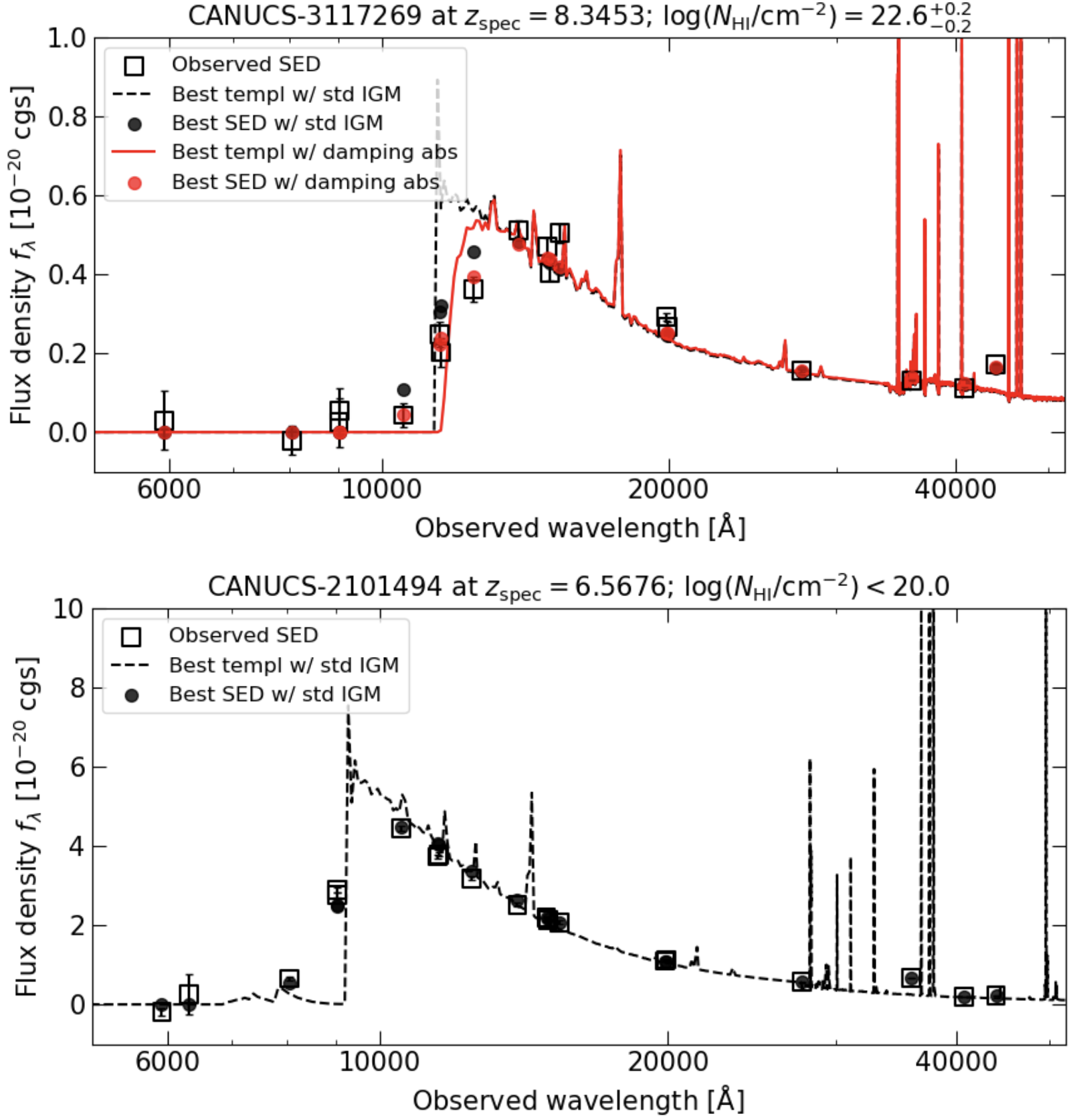}
    \caption{Example SEDs of the $N_{\rm H\textsc{i}, CGM}$ measurements in Section \ref{subsec:meas_DLA}.
    {\it Top}: CANUCS-3117269 is at $z_{\rm spec}=8.3453$, and the observed photometry (black squares) shows a flux deficit around the Lyman break as compared to the best-fitting template spectrum at $z_{\rm spec}$ (black dashed line; black circles are the in-band fluxes of the best template).
    A damping absorption with $\log(N_{\rm HI}/{\rm cm^{-2}})=22.6$ is required to reproduce the observed SED. Red curve shows the best template spectrum after the additional damping wing absorption, whose in-band fluxes are shown by red circles.
    {\it Bottom}: CANUCS-2101494 is at $z_{\rm spec}=6.5676$, and the observed photometry can be explained well without the additional damping wing absorption.
    The \hst/WFC3 IR F110W photometry is not shown for clarity.
    }
    \label{fig:SEDs}
\end{figure}

Figure \ref{fig:SEDs} shows two examples of $N_{\rm H\textsc{i}, CGM}$ measurements. The galaxy CANUCS-3117269 (top) is at $z_{\rm spec}=8.345$, and the observed SED (black squares) shows clear flux deficit in the spectral region just redward the Ly$\alpha$ line as compared to the best-fitting template spectrum (black curve, and black circles represent the in-band fluxes of  the best fit spectrum), which is an obvious signature of Ly$\alpha$ damping wing absorption.
A significant Ly$\alpha$ damping wing absorption is thus required to properly model the observed shape of SED, and $\log_{10}(N_{\rm H\textsc{i}, CGM}/{\rm cm^{-2}})=22.6\pm0.2$ gives the best-fit spectrum (red curve and red circles represent the in-band fluxes of the best-fit spectrum with CGM absorption).
On the other hand, galaxy CANUCS-2101494 (bottom panel) at $z_{\rm spec}=6.5676$, has no flux deficit around the break, and no significant CGM H{\sc i} absorption is required to reproduce the SED.

\subsection{Characterizing the $N_{\rm H\textsc{i}, CGM}$-$z$ relation}\label{subsec:NHI-z}

\begin{figure}[tb]
    \centering
    \includegraphics[width=0.95\linewidth]{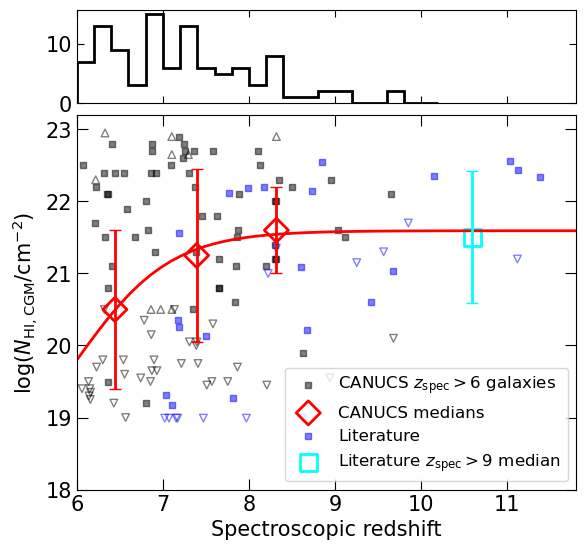}
    \caption{$N_{\rm H\textsc{i}, CGM}$ vs redshift relation.
    Black symbols denotes the $N_{\rm H\textsc{i}, CGM}$ measurements in this work from CANUCS $z_{\rm spec}>6$ sample. Black open upward/downward triangles represent lower/upper limits.
    The median value in each redshift bin are shown by red open diamonds.
    Blue symbols are $N_{\rm H\textsc{i}, CGM}$ measurements from literature \citep[][]{Carniani2024arXiv,D'Eugenio2024A&A,Hainline2024ApJ_b,Heintz2024Sci,Umeda2023arXiv}, and the cyan open square is the median value of the literature galaxies at $z_{\rm spec}>9$. Some of the literature data points at $z>12$ are not displayed in this figure.
    The red solid curve shows the best-fitting sigmoid parameterization of the $N_{\rm H\textsc{i}, CGM}$-$z$ relation. 
    }
    \label{fig:NHI_z}
\end{figure}

Gray symbols in Figure \ref{fig:NHI_z} show the individual measurements of $N_{\rm H\textsc{i}, CGM}$ for the 102 sample galaxies.
Based on these measurements, we next characterize the redshift evolution of $N_{\rm H\textsc{i}, CGM}$ at $z>6$. To this end, we compute the median redshift and $N_{\rm H\textsc{i}, CGM}$ of sample galaxies in each redshift bin ($z=6$--$7$, $z=7$--$8$, and $z>8$), and show them as red open diamonds in Figure \ref{fig:NHI_z}.
We also derive the median absolute deviations (MADs) of $N_{\rm H\textsc{i}, CGM}$, shown as the red error bars, to quantify the object-to-object scatter in each redshift bin.
In this calculation, we take into account the upper/lower limits by replacing them with their expected values computed assuming a top-hat probability distribution function confined within the 1-sigma upper/lower limits and the prior $N_{\rm H\textsc{i}, CGM}$ range of $\log_{10}( N_{\rm H\textsc{i}, CGM}/{\rm cm^{-2}})=[18.0, 23.0]$.
%The median $N_{\rm H\textsc{i}, CGM}$ values monotonically increase from $z\sim6$ to $z\sim8$, which is consistent with the picture of the cosmic reionization at this epoch.

Our sample from CANUCS observations does not include many galaxies at $z>9$ and so we do not have a good constraint on the redshift evolution of $N_{\rm H\textsc{i}, CGM}$ at $z>9$.
Therefore, we collect $N_{\rm H\textsc{i}, CGM}$ measurements at $z>9$ based on \jwst/NIRSpec observations from literature \citep[][]{Carniani2024arXiv,D'Eugenio2024A&A,Hainline2024ApJ_b,Heintz2024Sci,Umeda2023arXiv}, and they are plotted with blue symbols in Figure \ref{fig:NHI_z}. 
Here we only use galaxies in literature whose redshifts are determined by emission lines to mitigate the spec-$z$ uncertainties due to the damping absorption. We then compute the medians of the redshift and $N_{\rm H\textsc{i}, CGM}$ of literature galaxies at $z_{\rm spec}>9$, shown as cyan open square in Figure \ref{fig:NHI_z}.
The MAD is also computed and shown with the cyan error bar. The median $N_{\rm H\textsc{i}, CGM}$ value at $z_{\rm spec}>9$ galaxies from the literature is similar to that in the highest redshift bin from our CANUCS sample ($z\sim8.5$), which suggests almost no evolution of damping wing absorption at $z>8$.
This is consistent with \citet{Heintz2024arxiv}, who showed that the strength of the damping wing shape does not evolve significantly at $z>8$.
%Thi is consistent with the previous study \citet{Heintz2024arxiv}.

We then quantify the redshift evolution of $N_{\rm H\textsc{i}, CGM}$ with a simple analytical form that will allow us to use this $N_{\rm H\textsc{i}, CGM}$-$z$ relation to compute the median CGM optical depth $\tau_{\rm CGM}(\lambda_{\rm obs}, z_s)$ at a given redshift.
We fit a sigmoid function centered at $z=6$ to the median values (red and cyan open symbols in Figure \ref{fig:NHI_z}),
\begin{equation}\label{eqn:Sigmoid}
    \log_{10}[N_{\rm H\textsc{i}, CGM} (z_s)] = \frac{A}{1+\exp{[-a(z_s-6)]}} + C,
\end{equation}
and derive the empirical $N_{\rm H\textsc{i}, CGM}$-$z$ relation.
The best-fit parameters are $A=3.592$, $a=1.841$, and $C=18.001$, as shown with the red curve in Figure \ref{fig:NHI_z}.

\section{Results and discussions}\label{sec:results}

 \subsection{New transmission curves and photometric redshift improvements}\label{subsec:photz}

\begin{figure}[tb]
    \centering
    \includegraphics[width=0.95\linewidth]{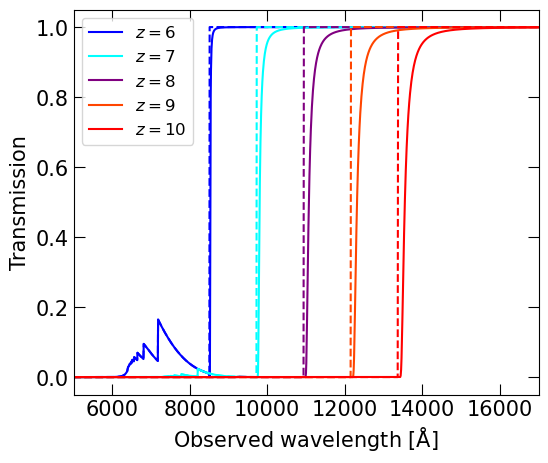}
    \caption{New IGM+CGM transmission curves at $z=6,7,8,9,10$. The dashed curves show the original IGM model by \citet{Inoue2014mn}, and the solid curve show the new transmission with damping absorption.
    }
    \label{fig:trans}
\end{figure}

\begin{figure*}[t]
    \centering
    \includegraphics[width=0.95\linewidth]{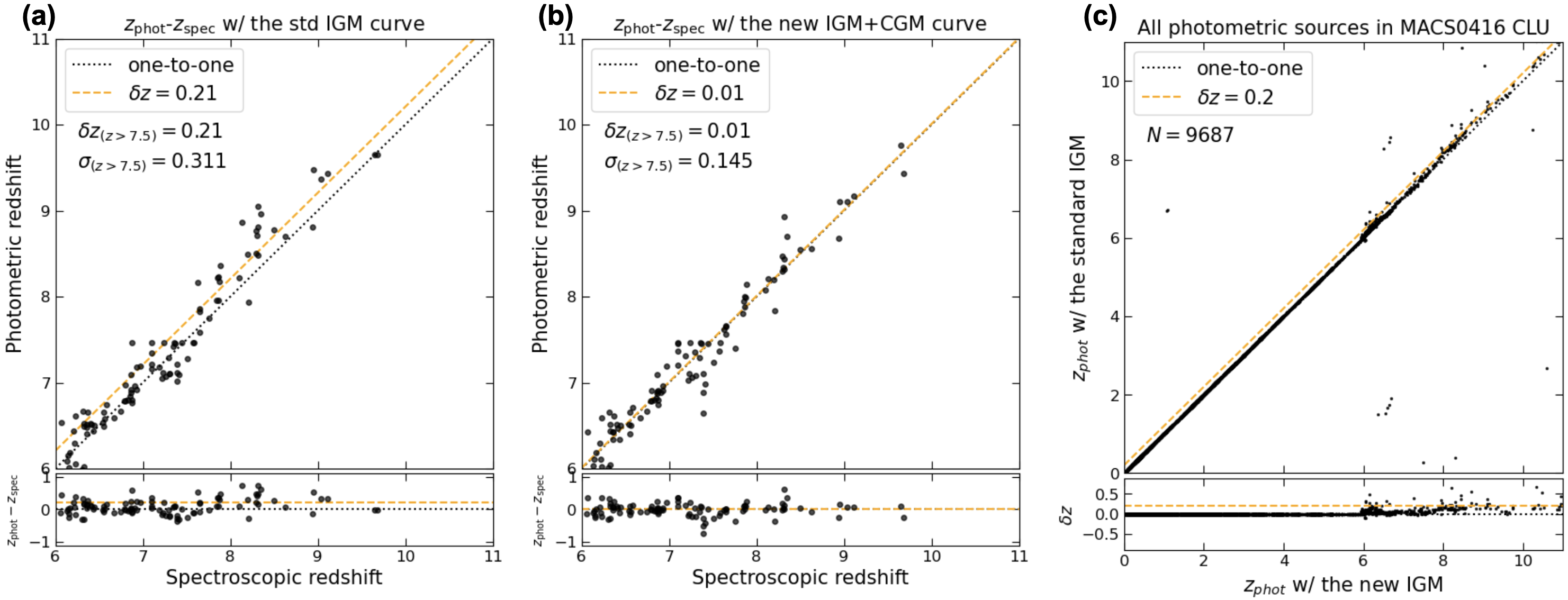}
    \caption{Photo-$z$ improvements with the new IGM+CGM model.
    \textbf{(a)} $z_{\rm phot}$-$z_{\rm spec}$ plot with the \citet{Inoue2014mn} IGM model. $z_{\rm phot}$ are systematically overestimated when the standard IGM transmission is assumed at $z_{\rm spec}>7.5$, and the median offset ($\delta z = z_{\rm phot} - z_{\rm spec}$) is 0.21 at the redshift range (orange dashed lines).
    \textbf{(b)} Same as panel (a) but with the new IGM+CGM transmission. The new transmission significantly improves $z_{\rm phot}$ estimations, and $z_{\rm phot}$ and $z_{\rm spec}$ agree well at all $z>6$.
    \textbf{(c)} The effect of the new IGM+CGM transmission on a catalog of all F277W ${\rm S/N}>5$ photometric sources. The new transmission has almost no effect on low-$z$ photo-$z$ estimations but corrects the bias by $\delta z\sim0.2$ at $z\gtrsim7.5$.
    }
    \label{fig:zphot_zspec}
\end{figure*}

Having the $N_{\rm H\textsc{i}, CGM}$-$z$ relation shown by the red curve in Figure \ref{fig:NHI_z}, we can compute the new IGM+CGM transmission curves at $z_{\rm spec}>6$.
Figure \ref{fig:trans} shows our new IGM+CGM transmission curve (solid curves), computed for five discrete redshifts, compared to the original IGM transmission curve of \citet[dashed curves]{Inoue2014mn}.
The damping absorption feature from the CGM H{\sc i} gas is apparent at higher redshift reaching a plateau at $z>9$, while the CGM contribution is negligible at $z=6$.

We stress that the Ly$\alpha$ emission line in the template spectra is attenuated with this additional CGM absorption.
In general, the inclusion of the Ly$\alpha$ emission line in the templates improves the photo-$z$ estimation in low-$z$ galaxies, but can also introduce  photo-$z$ overestimation at high-$z$ due to the low occurrence rate of Ly$\alpha$ emitters at these redshifts \citep[e.g.][]{Tang2024}.
Several studies of high-$z$ galaxies thus removed the Ly$\alpha$ emission line from the templates only at high-$z$ to partially alleviate the bias \citep[e.g.,][]{Willott2024ApJ,Hainline2024ApJ}.
However, with our new IGM+CGM transmission curves, the Ly$\alpha$ line in the template, even if it has a large equivalent width, is attenuated naturally by the CGM damping wing component that appears only at $z>6$. Consequently, when using the new IGM+CGM transmission curves, it is not necessary to implement a redshift-dependent Ly$\alpha$ exclusion from the template spectrum.
%We thus turn on the CGM component only at $z\ge6$, and use classical IGM transmission curve at $z<6$.

We next derive the photo-$z$s of our sample of 102 galaxies with this new IGM+CGM transmission curve and with the standard IGM transmission curve, and compare them to assess the improvement of photo-$z$ estimations with the new IGM+CGM transmission curve. The template-fitting code {\tt EAzY-py} is used to compute the photo-$z$. The same template set is used as in Section \ref{subsec:meas_DLA}, and no zero-point correction is applied.

Figure \ref{fig:zphot_zspec} shows the results.
When the standard IGM transmission curve is used, the photo-$z$ is systematically larger than the spectroscopic redshifts at $z\gtrsim7.5$, though they agree well at $6<z\lesssim7.5$ (panel (a)).
The median offset ($\delta z = z_{\rm phot} - z_{\rm spec}$) at $z>7.5$ is $\delta z_{z>7.5} = 0.21$, and the normalized median absolute deviation of $z_{\rm phot} - z_{\rm spec}$ is $\sigma_{z>7.5}=0.31$.
In contrast, when our new IGM+CGM transmission curve calibrated in this work is used, the systematic offset at $z>7.5$ is resolved while there is no effect at $6<z\lesssim7.5$. The photo-$z$ and spec-$z$ agree well at all $z>6$, including at the previously problematic $z>7.5$, where now $\delta z_{z>7.5} = 0.01$ and $\sigma_{z>7.5}=0.15$ (panel (b)).
This figure demonstrates that the new IGM+CGM transmission properly models the observed SEDs in high-$z$ galaxies and provides a good representation of typical high-$z$ galaxies' line-of-sight neutral hydrogen absorption.

To explore the effect of applying the new transmission curve on the photo-$z$ estimations over the full redshift range, we derive the photo-$z$ of the full photometric sample from one CANUCS field with the new IGM+CGM transmission and with the standard IGM transmission.
For this comparison, we use the CANUCS photometry catalog in the MACS0416 CLU field, and run {\tt EAzY-py} with the new IGM+CGM model and with the standard IGM model, for all photometric sources that are observed with all NIRCam filters available and have ${\rm S/N}>5$ in \jwst/NIRCam F277W filter.
In the IGM+CGM run, we turned off the CGM component $\tau_{\rm CGM}(\lambda_{\rm obs}, z_s)$ at $z<6$ because the CGM contribution is already negligible at $z=6$ (Figure \ref{fig:trans}) and the $N_{\rm H\textsc{i}, CGM}$-$z$ relation is calibrated at $z\ge6$.
Figure \ref{fig:zphot_zspec} (c) shows the comparison of photo-$z$ with and without the additional CGM absorption.
The figure shows that implementing the new IGM+CGM transmission has almost zero effect on photo-$z$ estimation at low-$z$, but corrects the bias of photo-$z$ at $z\gtrsim7$ by $\delta z \sim0.2$.

One might worry that a large number of low-$z$ Balmer break galaxies are (mis)identified as high-$z$ Lyman break galaxies with the new IGM+CGM transmission curve, since the new transmission makes the Lyman break smoother. 
We can see such a group of galaxies in Figure \ref{fig:zphot_zspec} (c), whose photo-$z$ used to be $z<3$ with the standard IGM but is $z>6$ with the new IGM+CGM curve.
But, this occurs only in eight sources of $\sim9700$ photometric sources ($\sim0.1\ \%$), thus the effect is not significant at all on the whole population. 
Conversely, those eight objects represent $\sim2\ \%$ of the $z>6$ sample, so even if they were really at low-$z$, they do not intrude a large contamination. 
We note that the small number of sources whose redshifts changed by a large amount are typically faint galaxies with a weak break in F090W
and a double-peaked photo-$z$ solution. Their photo-$z$ are thus inherently  ambiguous regardless of which transmission curve we adopt.  Without spectroscopy, their redshifts will remain uncertain.
%it is still possible that they are real high-$z$ faint galaxies. Since we do not have spectrosocpic redshift measurements of these galaxies, it is unclear which redshift solution is correct.

\subsection{Verification with other setups}

\begin{figure*}[t]
    \centering
    \includegraphics[width=0.95\linewidth]{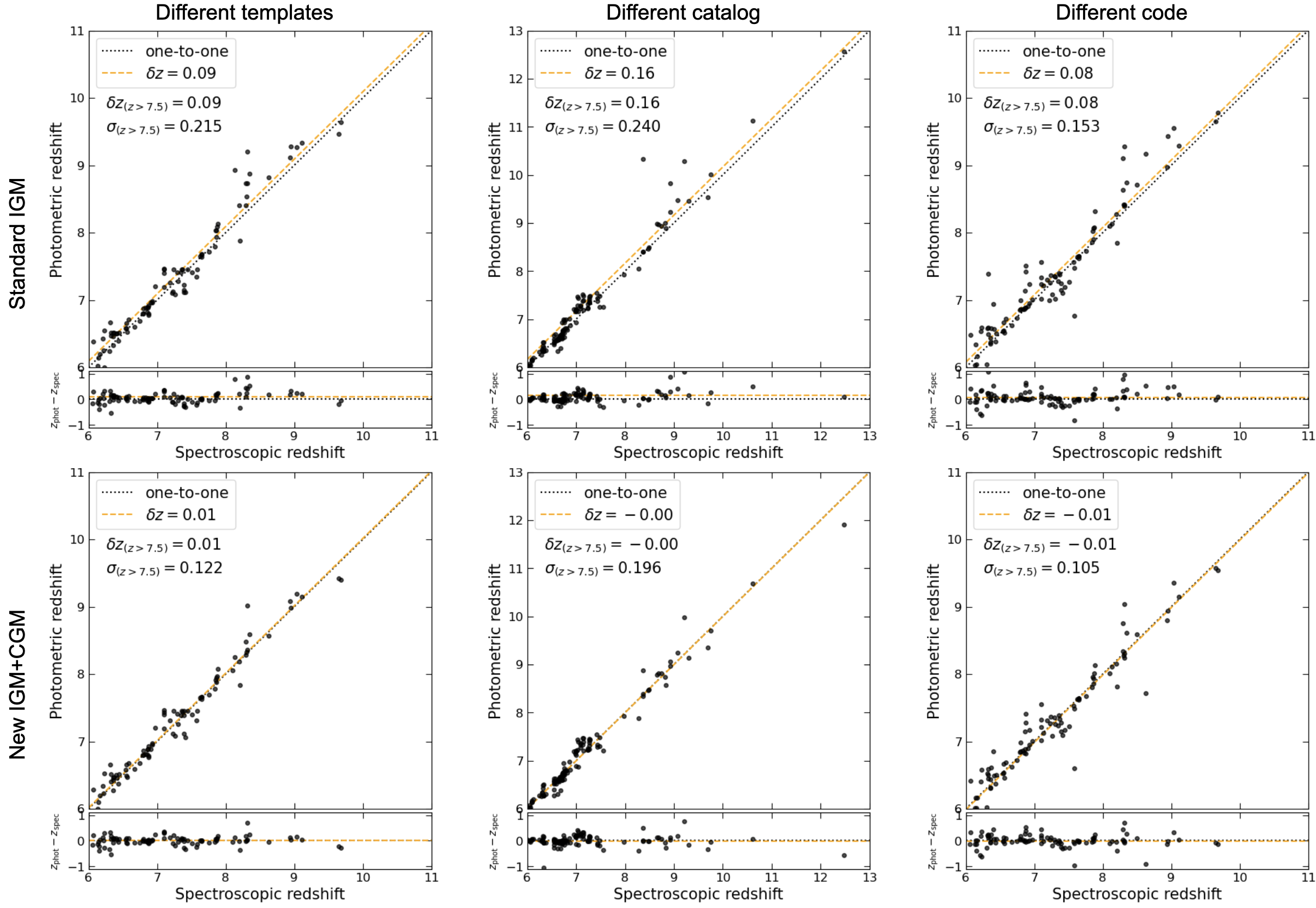}
    \caption{Verification of the new IGM+CGM transmission with different setups. The top row are $z_{\rm phot}$-$z_{\rm spec}$ plots with the standard \citet{Inoue2014mn} IGM model (similar to Figure \ref{fig:zphot_zspec} (a)), and the bottom row are those with the new IGM+CGM transmission (Figure \ref{fig:zphot_zspec} (b)). From left to right, comparisons with a different template set (left), a different photometric catalog by JADES (middle), and a different photo-$z$ code (right) are shown. The new transmission remarkably improves photo-$z$ estimations in all cases.
    }
    \label{fig:diff}
\end{figure*}

We showed in Sec.~\ref{subsec:photz} that the new transmission curve 
%composed of the intervening H{\sc i} gas in the line-of-sight IGM and dense H{\sc i} gas cloud in the CGM 
corrects the photo-$z$ bias at high-$z$ and represents the typical H{\sc i} gas properties of galaxies in the epoch of reionization.
However, these results are based on a specific template set (i.e., the custom templates described in Section \ref{subsec:meas_DLA}), a specific dataset (i.e., \jwst/NIRSpec sample from CANUCS observations), and a specific template-fitting code (i.e., {\tt EAzY-py}).
Although the photo-$z$ overestimations at $z>7$ when using the standard IGM transmission have been reported in several papers \citep[][]{Fujimoto2023ApJ,Helton2023arXiv,Finkelstein2024ApJ,Hainline2024ApJ,Willott2024ApJ}, it is worth investigating whether the new IGM+CGM curve works in different setups than the one we used in Sec.~\ref{subsec:photz}.
To this end, we compare the standard IGM with the new IGM+CGM as in Section \ref{subsec:photz} for three different setups.

The first test is with a different template set.
Different studies in the literature often use different template sets and it is thus important to see if the new transmission curve works with other template sets.
For this test, we use the {\tt sfhz} template set from {\tt EAzY-py}\footnote{\url{https://github.com/gbrammer/eazy-photoz/tree/master/templates/sfhz}}, and compute the photo-$z$ for our 102 NIRSpec galaxies at $z_{\rm spec}>6$, with and without the CGM absorption as we did in Section \ref{subsec:photz}.
The left column in Figure \ref{fig:diff} show the results.
Even when a different template set is used, a similar photo-$z$ bias can be seen with the standard IGM model (top left panel; $\delta z_{z>7.5} = 0.09$), but this bias is resolved with the new IGM+CGM model (bottom left panel; $\delta z_{z>7.5} = 0.01$).

The second test is with a different data set.
It is essential to see the validity of the new transmission with different photometric catalogs, particularly because doing so ensures that the test sample and the training sample used to constrain the parameters of the model (Eq.~\ref{eqn:Sigmoid}) are independent of each other.
To this end, we utilize \jwst\ observations by the JWST Advanced Extragalactic Survey (JADES) program \citep[][]{Eisenstein2023arXiv_a}.
We use the v2 photometric catalogs in the GOODS-S deep field \citep[][]{Eisenstein2023arXiv_b} and v1 photometric catalogs in the GOODS-N field \citep[][]{D'Eugenio2024arXiv}.
For the spectroscopic redshifts, we use the v1.1 NIRSpec catalogs in the GOODS-S and GOODS-N fields \citep{Bunker2023arXiv,D'Eugenio2024arXiv}.
With these JADES datasets, we make a comparable sample of 128 galaxies at $z_{\rm spec}>6$ in the GOODS-S and GOODS-N fields by requiring a secure redshift of $z_{\rm spec}>6$ and reliable photometry in multiple NIRCam filters.
We then use our standard template set to run {\tt EAzY-py} on the 0.3$^{\prime\prime}$-diameter aperture photometry given by the JADES catalog, with the standard IGM transmission and the new IGM+CGM transmission to compute the photo-$z$ in each case. The Figure \ref{fig:diff} middle column shows the results.
\citet{Hainline2024ApJ} found that the photo-$z$ overpredict the spec-$z$ in general at $z>7$ using the JADES dataset ($\delta z$ = 0.26), and we get a similar offset when the standard IGM curve is used (top middle panel; $\delta z_{z>7.5} = 0.16$).
However, this offset disappears when the new IGM+CGM transmission is used instead (bottom middle panel; $\delta z_{z>7.5} = 0.00$).

The final test is with a different photo-$z$ template-fitting code.
It is critical to test if the new transmission curve is valid when a different template-fitting code is utilized to compute the photo-$z$.
For this test, we use the sample of 102 CANUCS spec-$z$ galaxies, and run {\tt Phosphoros}\footnote{\url{https://anaconda.org/ astrorama/phosphoros}} \citep[][Paltani et al.\ in prep.]{Desprez2020A&A} to compute the photo-$z$.
The templates used in {\tt Phosphoros} to fit the source SEDs are those from the \citet{Ilbert2013} COSMOS library complemented with set 1 and 4 from \citet{Larson2023ApJ}. Source intrinsic extinction is handled in the same way as in \citet{Desprez2023} for the COSMOS templates, and the configuration presented in \citet{Desprez2024MNRAS} is used for the Larson templates. Emission lines (Balmer lines, [O{\sc iii}] and [O{\sc ii}]) are added to the COSMOS templates as Gaussian lines with FWHM of $200\,{\rm km\,s^{-1}}$ according to the recipe implemented in \texttt{Phosphoros} \citep{Paltani2024}, and Ly$\alpha$ emission is included with eight times the flux of H$\alpha$. No magnitude prior is applied to the fit, and the redshift probability distribution function $P(z)$ is estimated from $z=0$--$20$ with a $\delta z = 0.01$ step size. The median values of the $P(z)$ is adopted as the photo-$z$ point value for plotting. %The standard IGM is set the \citet{Inoue2014mn} one. 
The right column of Figure \ref{fig:diff} shows the results.
As in the earlier tests, the photo-$z$ offset at $z>7.5$ that is seen with the standard IGM transmission (top right panel; $\delta z_{z>7.5} = 0.08$) is significantly improved with the new IGM+CGM transmission (bottom right panel; $\delta z_{z>7.5} = -0.01$).

As shown in this subsection, the photo-$z$ are systematically overestimated when the standard IGM transmission is assumed at $z>7.5$, regardless of the dataset, the template set, or the template-fitting code.  However, in each of these cases the overestimation is resolved with our new IGM+CGM transmission curve that includes the additional damping wing absorption.
This demonstrates universal applicability of the new IGM+CGM transmission curve.

\subsection{Other implications}\label{subsec:implication}

Although the main goal of this paper is to model the damping absorption feature to improve photo-$z$ estimations at $z>7$, the $N_{\rm H\textsc{i}, CGM}$-$z$ relation shown in Figure \ref{fig:NHI_z} also gives interesting hints about the properties of EoR galaxies. We consider two possible scenarios. 

If we assume that the smoothed shape of the Lyman break at $z>7$ observed in this paper is primarily due to the dense neutral gas clouds proximate to galaxies, then Figure \ref{fig:NHI_z} suggests that galaxies reside within denser H{\sc i} gas (i.e., more neutral environments) from $z\sim6$ up to $z\sim8$, but there is no significant evolution above $z\sim8$.
This is consistent with the idea that the reionization process mostly takes place after $z\sim8$, favoring a more rapid and late reionization scenario \citep[e.g.,][]{Ishigaki2018ApJ,Naidu2020ApJ} over a more extended and early transition \citep[e.g.,][]{Finkelstein2019ApJ}. Similar rapid reionization process has been indicated by recent studies of damping wing absorption based on \jwst/NIRSpec spectroscopic observations \citep[e.g.,][]{Umeda2023arXiv,Heintz2024arxiv}.

Alternatively, if the Lyman break smoothness at $z>7$ in our sample galaxies is mainly due to increased prominence of nebular continuum in the rest-frame UV spectra, the growing smoothness shown in Figure \ref{fig:NHI_z} may suggest that sub-$L^\star$ galaxies at $z>7$ possess very extreme nature in general. The two-photon emission of the nebular continuum can dominate over stellar emission throughout the rest UV rangeonly when the galaxy has a remarkably top-heavy IMF \citep[e.g.,][]{Cameron2024mn,Katz2024arXiv,Mowla2024arXiv}.
Therefore, the redshift evolution in Figure \ref{fig:NHI_z} may be due to non-universal IMFs in sub-$L^\star$ galaxies in the EoR.

These possible physical origins of the smoothed Lyman break have different spectral signature, though they are degenerated in the photo-$z$ estimations where a limited number of photometric data points are available around the break \citep[even NIRSpec/{\tt PRISM} can lack the spectral resolution; e.g.,][]{Huberty2025arXiv}. Consequently, we do not intend to disentangle the physical origin in this paper.
Future deep and high resolution spectroscopic  observations are  needed to reveal the physical origin of the increasing smoothness of the Lyman break in the EoR.

%Another implication can be found from the object-to-object scatter of $N_{\rm H\textsc{i}, CGM}$ values and relates to the patchiness of the reionization process. Here, we quantified the amount of $N_{\rm H\textsc{i}, CGM}$ scatter in each redshift bin by deriving the median absolute deviation (MAD).  The scatter (MAD) values are shown as red error bars in Figure \ref{fig:NHI_z} and can be seen to increase as reionization proceeds ($\sigma_{z=6-7}=1.0$ dex, $\sigma_{z=7-8}=1.0$ dex, $\sigma_{z=8-10}=0.7$ dex). This increasing scatter is consistent with a patchy cosmic reionization process, in agreement with previous theoretical \citep[e.g.,][]{Barkana2001PhR} and observational studies \citep[e.g.,][]{Becker2015MNRAS}.

%\section{Caveats}
\subsection{Caveats: the $N_{\rm H\textsc{i}, CGM}$ values from photometry}
In this work, we model the damping wing absorption feature based on photometry and assuming the classical IGM transmission curve \citep{Inoue2014mn}, to measure the CGM H{\sc i} column density $N_{\rm H\textsc{i}, CGM}$ (see Figure \ref{fig:NHI_z}).
Although the good agreement between photo-$z$ and spec-$z$ shows that the net optical depth $\tau_{\rm tot}(\lambda_{\rm obs},z_s)$ obtained in this work represents typical galaxies' SEDs in the EoR, the $N_{\rm H\textsc{i}, CGM}$ values themselves could have systematics, and the physical origins of the smoothed shape of the break are impossible to interpret at this point.
As discussed in the literature \citep[e.g.,][]{Umeda2023arXiv,Heintz2024arxiv,Keating2024MNRAS}, the damping wing absorption can be due to dense H{\sc i} gas clouds in the CGM but also due to a high neutral fraction and/or high density of the line-of-sight IGM gas.
In some extreme case, as mentioned in Section \ref{subsec:implication}, two-photon emission of the nebular continuum can dominate in the rest-frame UV spectra and mimic the Ly$\alpha$ damping wing absorption of a dense neutral hydrogen cloud \citep[e.g.,][]{Cameron2024mn,Katz2024arXiv}.
Another possible systematic is due to sample selection; as mentioned in Section \ref{subsec:meas_DLA}, we do not require the detection of the rest-UV continuum in the spectrum in our sample selection, and our galaxy sample contains a large number of faint emission-line-only galaxies, contrary to other studies of $N_{\rm H\textsc{i}, CGM}$ at high-$z$ which commonly require high S/N detection of the rest-UV continuum in the spectra.
The differences in sample selection and methodology may also introduce systematic differences in $N_{\rm H\textsc{i}, CGM}$.
%The precise modeling involving both IGM and CGM contribution to the damping wing absorption requires high-resolution spectroscopic observation (even \jwst/NIRSpec PRISM observation is not enough).
However, the overall redshift evolution of $N_{\rm H\textsc{i}, CGM}$ shown in Figure \ref{fig:NHI_z} is reasonable and consistent with previous studies based on \jwst/NIRSpec observations \citep[][]{Umeda2023arXiv,Heintz2024Sci,Heintz2024arxiv}, and thus the relation itself is sufficient for current use of photo-$z$ determination. 
%but we suggest caution when using the $N_{\rm H\textsc{i}, CGM}$-$z$ relation from this work.
%Moreover, and despite the fact that some assumptions that go into our model may not be correct (such as the nature of the gas that we call "CGM" throughout this work), the ultimate test of the usefulness of our new transmission curve is in its ability to remove the bias commonly seen in photo-$z$ at high-$z$. This is where the usefulness and validity of the new transmission curve truly lies. 

\iffalse
\subsection{Validity at the highest redshift end}
Another caveats is the validity of the new IGM+CGM transmission at the highest redshift end of $z>11$.
There are not enough constraint at $z>11$ yet, and it is not clear if the new transmission works well enough at the highest redshift end.
Further investigation will be needed when a larger sample of galaxies at these highest redshift is available.
\fi

\section{Summary}\label{sec:summary}
In this letter, we present a novel analytic model for the attenuation by the line-of-sight neutral hydrogen gas in high-$z$ galaxies.
After the launch of \jwst, it has been shown that the photo-$z$ are systematically overestimated, particularly at $z\gtrsim7$, and this bias is deemed to originate from the galaxy templates being too blue in the spectral region just redward of the Lyman break.
We exploit \jwst\ observations by the CANUCS program, and model the Ly$\alpha$ damping wing absorption feature by considering dense H{\sc i} gas clouds locally associated with high-$z$ galaxies.
We measure the proximate H{\sc i} gas column density causing the Ly$\alpha$ damping wing based on the NIRCam photometry in $z_{\rm spec}>6$ galaxies, and derive an analytical model for the $N_{\rm H\textsc{i}, CGM}$-$z$ relation by fitting a sigmoid function to the median redshift evolution of $N_{\rm H\textsc{i}, CGM}$-$z$ from $z_{\rm spec}=6$ to $10$ (Figure \ref{fig:NHI_z}).
This $N_{\rm H\textsc{i}, CGM}$-$z$ relation is then used to compute the optical depth of the proximate H{\sc i} clouds contributing the Ly$\alpha$ damping wing absorption ($\tau_{\rm CGM}(\lambda_{\rm obs},z_s)$) and the net optical depth ($\tau_{\rm tot}(\lambda_{\rm obs},z_s) = \tau_{\rm IGM}(\lambda_{\rm obs},z_s) + \tau_{\rm CGM}(\lambda_{\rm obs},z_s)$) at a given redshift $z_s$.
The new IGM+CGM transmission curve (Figure \ref{fig:trans}) includes strong damping wing absorption at $z\gtrsim7$, and it significantly improves photo-$z$ estimations at $z>7$ regardless of the choice of data set, template set, or even template-fitting code (Figure \ref{fig:zphot_zspec} and \ref{fig:diff}).
Also, adopting new IGM+CGM transmission at $z>6$ has almost no effect on the photo-$z$ estimations for low-$z$ galaxies (Figure \ref{fig:zphot_zspec} (c)).
The new CGM component is modeled independently of the commonly used IGM transmission curve \citep[][]{Inoue2014mn}, and thus the new IGM+CGM transmission can be easily implemented in any existing template-fitting codes simply by computing the CGM component transmission separately ($T_{\rm CGM} = \exp[-\tau_{\rm CGM}(\lambda_{\rm obs},z_s)]$) and applying the additional CGM transmission on the top of the IGM transmission used in that code.
This new IGM+CGM transmission curve is available in the latest versions of the {\tt EAzY-py} package and {\tt Phosphoros} template-fitting codes.
The proposed modeling of the Ly$\alpha$ damping wing absorption is also flexible and its calibration can be further improved by adjusting the three parameters in Eq. (\ref{eqn:Sigmoid}), particularly when a larger sample of spectroscopically confirmed $z>10$ galaxies is available.

\section*{Acknowledgements}

We thank the anonymous referee for the useful suggestions that improved the quality of this paper.
This research was enabled by grant 18JWST-GTO1 from the Canadian Space Agency, and Discovery Grant funding to MS and AM from the Natural Sciences and Engineering Research Council of Canada. 
This research used the Canadian Advanced Network For Astronomy Research (CANFAR) operated in partnership by the Canadian Astronomy Data Centre and The Digital Research Alliance of Canada with support from the National Research Council of Canada the Canadian Space Agency, CANARIE and the Canadian Foundation for Innovation.
YA is supported by a Research Fellowship for Young Scientists from the Japan Society of the Promotion of Science (JSPS) and by the JSPS International Leading Research (ILR) project (KAKENHI Grant Number JP22K21349).
MB acknowledges support from the ERC Grant FIRSTLIGHT, from the Slovenian national research agency ARIS through grants N1-0238, P1-0188, and the program HST-GO-16667, provided through a grant from the STScI under NASA contract NAS5-26555.

\section*{Data availability}
Data presented in this paper will be distributed upon  request. Also, the new IGM+CGM transmission curve is implemented in the latest version of the two template fitting codes ({\tt EAzY-py v0.8} and {\tt Phosphoros  v2.0.2}).
The CANUCS DOI is \href{https://doi.org/10.17909/ph4n-6n76}{doi: 10.17909/ph4n-6n76}.


\bibliography{sample631}{}
\bibliographystyle{aasjournal}



\end{document}